\documentclass[doublecol]{epl2} 
\usepackage{float}
\usepackage{dirtytalk}
\usepackage{textcomp}
\usepackage{hyperref}
\title{First neutron data recorded at the V20 Test Instrument utilizing prototype chopper systems and beam monitors planned for ESS}
\shorttitle{ESS Chopper and Detector tests at the V20 Test Instrument, HZB} 

\author{ V. Maulerova \inst{1,}   \inst{2} \and J. Nilsson \inst{1} \and M. Olsson \inst{1} \and S. Alcock \inst{1} \and A. Quintanilla \inst{1} \and D. Zielinski \inst{1} \and A. Mukai \inst{1} \and F. Issa \inst{1} \and S. Kolya \inst{1} \and W. Smith \inst{3} \and D. Broderick \inst{1} \and K. L\"{o}ki \inst{1} \and  J. Sparger \inst{1} \and R. Woracek \inst{1} \and J. Taylor\inst{1} \and T. Richter \inst{1} \and R. Hall-Wilton \inst{1,} \inst{4} \and O. Kirstein \inst{1,} \inst{5} \and N. Tsapatsaris \inst{1} \thanks{Corresponding author: \email{nikolaos.tsapatsaris@esss.se}} }
 
\shortauthor{ V. Maulerova \etal}

 \institute{                    
  \inst{1} European Spallation Source ERIC (ESS) - P.O. Box 176, SE-22100 Lund, Sweden.\\
  \inst{2} Lund University, Deparment of Physics, Nuclear Physics-P.O Box 118, SE-221 00, Lund, Sweden \\
    \inst{3} Science and Technology Facilities Council (STFC), Daresbury Laboratory, UK \\
  \inst{4} Università degli Studi di Milano-Bicocca, Piazza della Scienza 3, 20126 Milano, Italy\\
  \inst{5} University of Newcastle, School of Mechanical Engineering,  NSW, Australia \\

}
\pacs{29.85.Ca}{Data acquisition and sorting}
\pacs{29.25.Dz}{Neutron sources}
\pacs{29.27.Fh}{Beam characteristics}

\abstract{The main objective of the European Spallation Source (ESS) is to perform consistently high impact neutron scattering science using the highest neutron flux of any facility in its class. This ambition is naturally accompanied by operational challenges related to the vertical integration of the instruments{\textquotesingle} hardware and software shared between collaborators from 13 European countries.
The present work will detail integration tests performed at the V20 Test Instrument at Helmholtz-Zentrum Berlin in Germany.
An ESS chopper prototype was successfully controlled through the ESS Chopper Integration Controller (CHIC) and Experimental Physics and Industrial Control System (EPICS). 
Neutron data were successfully collected and time stamped using neutron beam monitors, the ESS prototype detector readout electronics and the Data Management Software Center (DMSC) data acquisition (DAQ) software. Chopper rotation events were timestamped and the chopper disk showed excellent stability and neutron absorption characteristics. The test resulted in (i) the expected time-of-flight (TOF) response, (ii) collected data that compared well with instrument reference data and (iii) revealing the diagnostic power of the performed integration tests. The results of this integration test validate the ESS chopper and detector DAQ architecture choices.
}
\begin{document}

\maketitle
\section{Introduction}
The European Spallation Source, a European Research Infrastructure Consortium (ERIC), is a multi-disciplinary research facility based on the world\textquotesingle s most powerful neutron source with a vision to enable scientific breakthroughs in research related to materials, energy, health and the environment, and address some of the most important societal challenges of our time \cite{TDRbook,TechnicalDesignReport, Garoby_2018}. The ESS will reach higher flux and longer pulse length compared to other neutron research facilities. In order to meet the needs of the diverse scientific areas, 16 neutron scattering instruments are currently foreseen in the baseline instrument suite. It is expected that the instruments, spanning over 3 instrument halls, requiring in total ca. 150 choppers and complex neutron detectors amongst other components, will introduce operational and technical challenges not previously seen in neutron scattering facilities. These include the software for high data rate acquisition, analysis, visualization and archiving as well as maintenance and integration of the state-of-the-art hardware provided by in-kind collaborators from 13 European countries. 

Unlike reactor sources and short pulsed spallation facilities, virtually all instruments at ESS will be equipped with chopper systems \cite{TechnicalDesignReport}. 
Neutron choppers are devices designed to periodically interrupt the neutron beam for a well defined duration. A neutron chopper operates as a form of \say{high speed} mechanical switch or filter for neutrons \cite{First_choppers}. It has two stable operating states: Open and Closed, and Transition states (opening/closing). Individually or in groups the neutron choppers can be configured to perform a number of functions to condition neutron beams for scientific uses. 
Depending on the type of instrument, a chopper system can be configured to produce the needed distribution of neutrons as a function of time.

Neutron detectors are devices collecting the charge induced by the neutrons via neutron interaction in a specially designed converter material\cite{Knoll}.
There is a large variety of neutron detectors that will be used at ESS \cite{multiblade,multigrid,gdgem,MaPMT,CDT,Kirstein2014,HENSKE2012151}. 
\par
In this work proportional wire chamber type Beam Monitors (BMs) were used \cite{CharacterizationBeamMonitors,ORDELA,Mirrotron} in order to demonstrate the vertical integration. With the help of read-out electronics and software \cite{4.1}, BMs can be used to record the intensity of the neutron beam as a function of time. Beam monitors are an essential component of the instrument suite and are vital input for reliable scientific investigations and functioning of the instrument. They are invaluable in monitoring the facility's performance and verifying satisfactory operation as well as helping diagnose problems at a neutron instrument. 
\par
These type of BMs are not designed to detect the entire neutron flux; on the contrary, the aim is to leave the beam undisturbed. They can be used to obtain the neutron energy distribution via the concept of the time-of-flight \cite{TOForigin}. Placing two BMs at known positions of the neutron path allows to record the arrival time of a sample of the ensemble of the neutrons passing through the first BM and second BM respectively. This information results in an indirect measurement of the energy of the neutrons, as shown in Eq.~\ref{eq:tof} where $\lambda$ is the wavelength of the neutron, $h$ is the Planck's constant, $t_2,t_1$ the time stamps from the BMs, $m$ the mass of the neutron and $L$ the distance between the two BMs: 
\begin{equation}
\lambda=\frac{h(t_2-t_1)}{mL}.
\label{eq:tof}
\end{equation}
This relation is fundamental to the operation of neutron time-of-flight instruments.

\begin{figure}[h]
\onefigure[width=8.5cm]{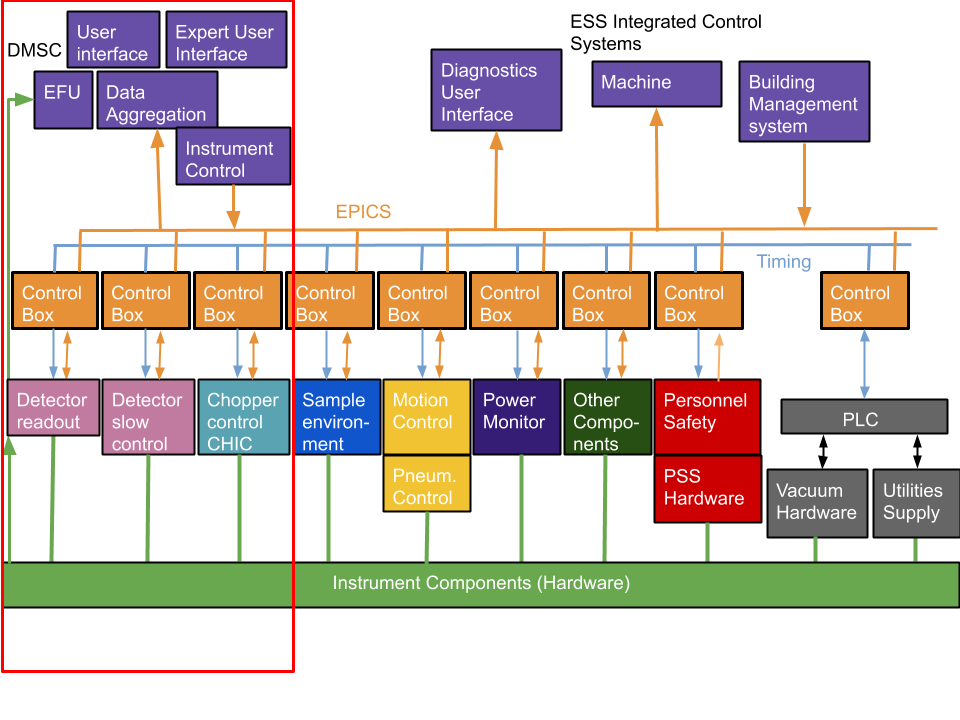}
\caption{ESS Controls and Readout System Architecture \cite{ESSVerticalIntegration}.}
\label{fig.Architecture}
\end{figure}

In order to efficiently employ time-of-flight neutron detectors, including BMs, as well as other devices that require accurate timing and synchronization to the pulsed neutron source, such as chopper systems, ESS has designed a modular controls and read-out architecture (Fig.~\ref{fig.Architecture}).
Instrument components are located at the lowest level. Through standardized interfaces (detector read-out, detector slow control, chopper control, etc.), these layers become exposed to the Experimental Physics and Industrial Control System (EPICS, \cite{EPICS}) backbone through Control Boxes (CB server). CBs fuse slow control information, time referencing and time stamping functionality. The timing system is centralized across the entire facility, providing a documented resolution of 11.357 ns based on a 88.0525 Mhz master clock \cite{ESSVerticalIntegration}. By taking any multiple of the master clock, each device can be given its own customized clock. For example the accelerator produces 14 bunches of protons per second, i.e. the repetition rate of the ESS source \cite{TechnicalDesignReport}. 
\par 
The top layer served by EPICS are the scientific user interfaces and ancillary functionalities provided by Data Management and Software Centre (DMSC). These final but important layers are allowing the data acquisition \cite{Mukai_2018}, the event formation \cite{Christensen_2018}, the instrument control, the data reduction \cite{Mantid}, monitoring, scientific data analysis and data archiving. 
\par
For the purpose of testing novel neutron methods as well as hardware and software solutions that are relevant for future instruments, ESS is operating a dedicated neutron time-of-flight test beamline (the V20 Test Instrument) at Helmholtz-Zentrum Berlin.

The main aim of the documented herein test, from now on referred to as the HZB test, is to demonstrate functioning vertical integration of (i) the Chopper Integration Controller (CHIC) \cite{CHIC}-EPICS control on a fully functional chopper prototype (ESS Chopper Prototype), (ii) time stamping and referencing in order to assess chopper phase and neutron monitor stability compared to each other and (iii) the detector read-out with the help of the DMSC data agreggation and file writer. The ESS functions utilized by the HZB test are shown in the Fig.~\ref{fig.Architecture} marked by the red square.


\section{Implementation of the test system}
\subsection{Experimental setup}	
\begin{figure}[h]
\onefigure[width=8.5cm]{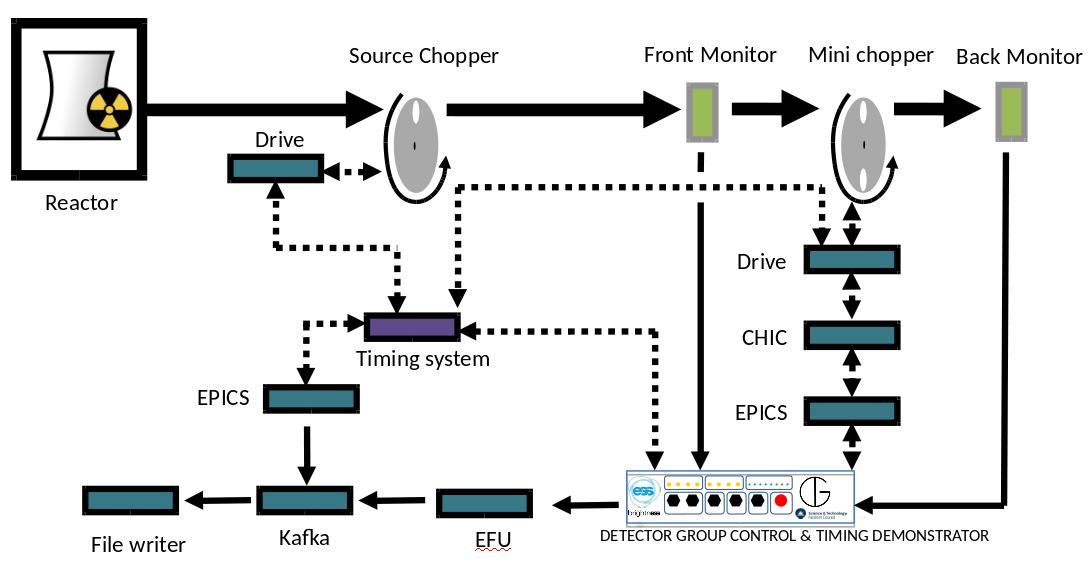}
\caption{Control system architecture.}
\label{fig.NICOS}
\end{figure}
\begin{figure*}[h]
\begin{center}
\includegraphics[width=\textwidth]{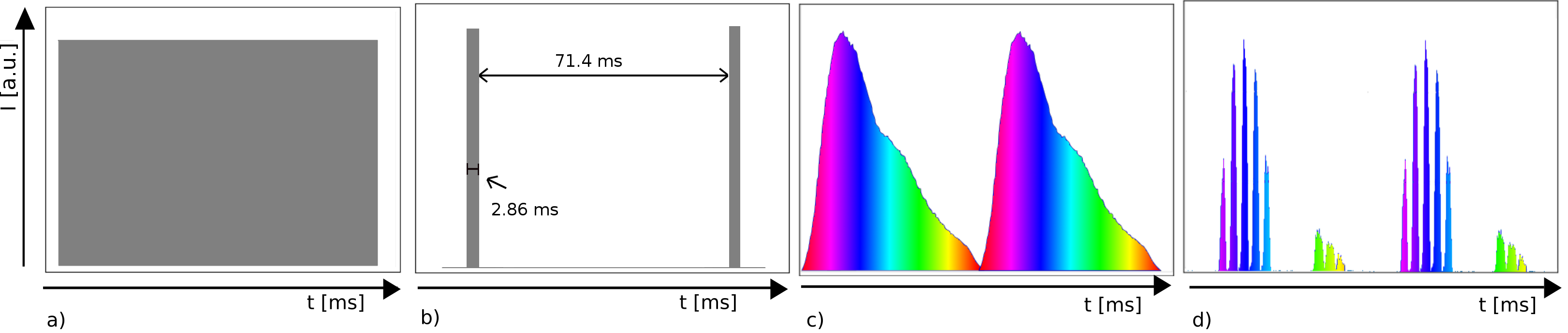}
\caption{Cartoon demonstrating the evolution of the signal in the HZB experimental setup: a)~the incoming neutron counts from the moderator, b)~the signal after the Source Pulse Choppers (with period 71.4 ms)  c)~the signal just before entering the ESS Chopper Prototype (smeared due to the neutron velocity) d) the signal right after the ESS Chopper Prototype. The y-axis (I) in all cases represents intensity in arbitrary units. The figures are conceptual; the x axis is not to scale.}
\label{fig.ChopperFunction}
\end{center}
\end{figure*}

 
The control system architecture utilized in the HZB test, performed at the Helmholtz Zentrum Berlin research reactor using the V20 Test Instrument \cite{WORACEK,STROBL201374}, is shown in detail in Fig.~\ref{fig.NICOS}.
The reactor provides a continuous stream of neutrons that are moderated by the cold neutron source and transported by neutron guides to the first chopper set (Source Pulse Choppers), see (Fig.~\ref{fig.ChopperFunction}a). The beam of neutrons is shaped by the Source Pulse Choppers. The Source Pulse Choppers are used to produce neutron pulses of approx. 2.86 ms length at a repetition rate of 71.4 ms, which mimic the behavior of the ESS neutron source. \cite{STROBL201374}. The pulse obtained right after the Source Pulse Choppers is shown in Fig.~\ref{fig.ChopperFunction}b (compare to experimental data in \cite{WORACEK, FORSTER2018298}).
Due to the different neutron velocities contained within the neutron pulse right after the Source Pulse Choppers, a spread of the pulse is observed before reaching the ESS Chopper Prototype, see Fig.~\ref{fig.ChopperFunction}c.
 Finally, the ESS Chopper Prototype transmits neutrons of selected energies according to the openings and phase of the disk with respect to the Source Pulse Choppers (Fig.~\ref{fig.ChopperFunction}d). The collection of neutrons after the ESS Chopper Prototype allows the verification of the chopper, detector, data acquisition and timing reference and timestamping system.
\subsection{Chopper system}
\begin{figure}[h]
\onefigure[width=6.5cm]{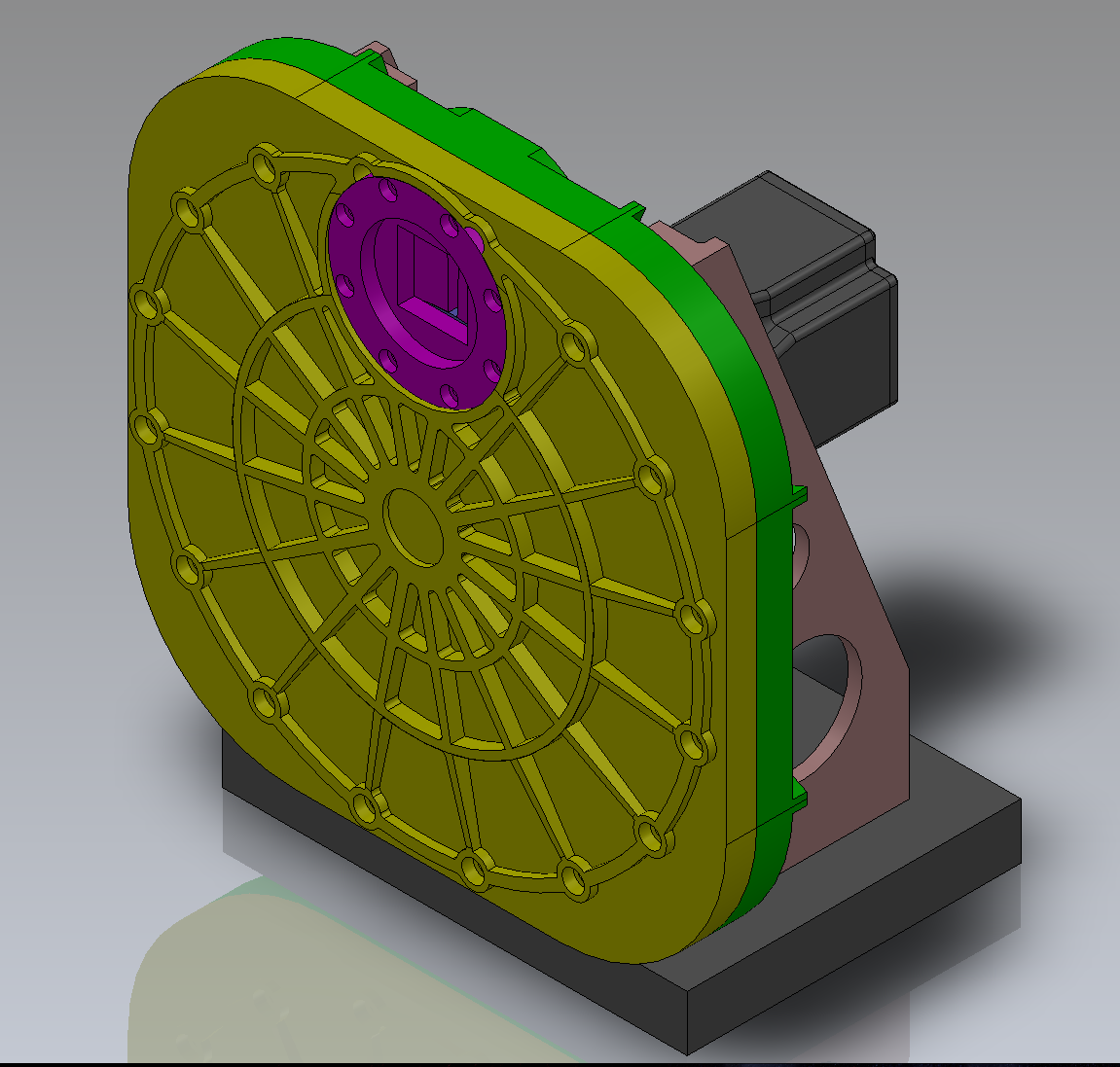}
\caption{CAD Drawing of the ESS Chopper Prototype prototype by Mastercam CAD software \cite{MasterCam}.}
\label{fig.minichopper}
\end{figure}
The configuration of the V20 chopper system used in these tests is composed of the Source Pulse Choppers, the Wavelength Band Choppers and the ESS Chopper Prototype. 
The V20 Wavelength Band Choppers were employed in order to prevent overlap between the slowest and fastest neutrons of two consecutive pulses.
The Source Pulse Choppers and Wavelength Band Choppers are synchronized to the the timing system via a TTL output signal and monitored only in terms of speed and phase through the timing system.
The ESS Chopper Prototype (Fig.~\ref{fig.minichopper}) contains a disk with an outer diameter of 175~mm and was operating at a frequency of 14~Hz. The size of the beam opening on the ESS Chopper Prototype is 25x25 mm. The disc is made of aluminium alloy (6082 T6/T651), coated by an epoxy/B$_4$C layer on both sides with the total thickness of 3.5~mm.
The ESS Chopper Prototype dimensions are a factor of 4 smaller than the choppers expected at ESS, but its functionality is completely identical to normal sized choppers.
\par
\begin{figure}[h]
\onefigure[width=6.5cm]{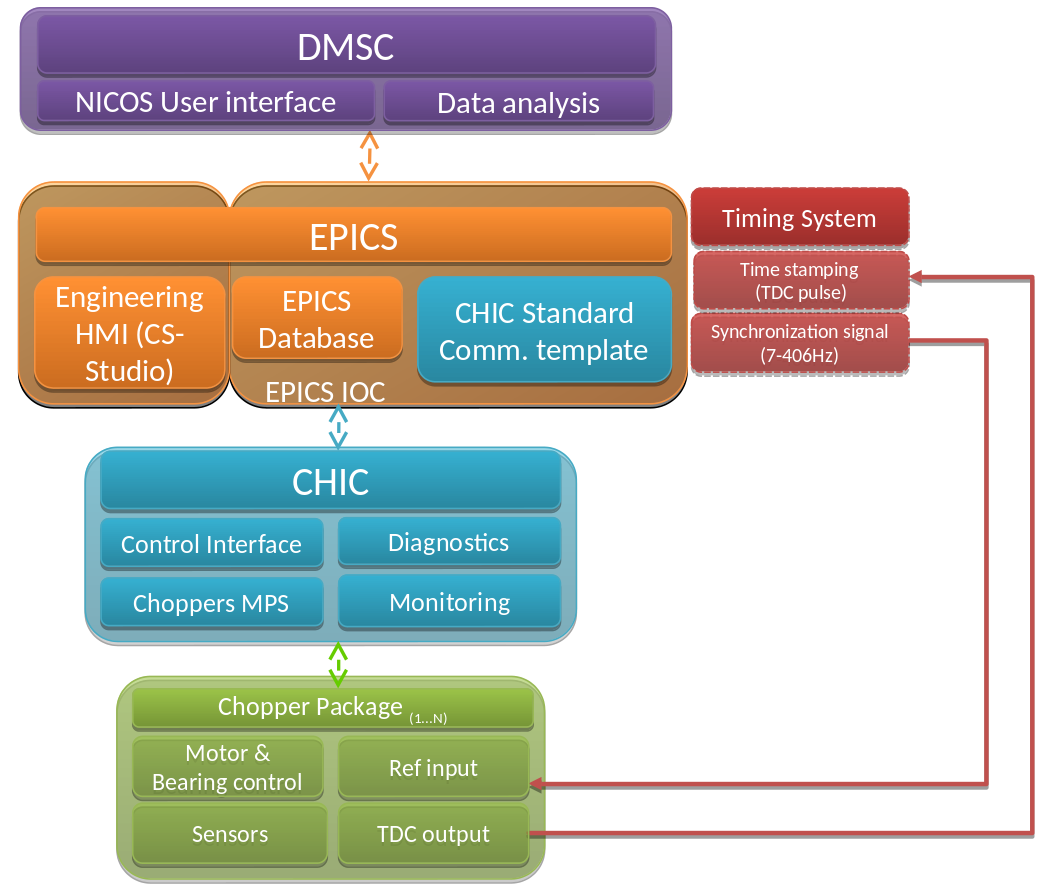}
\caption{ESS Chopper Prototype detailed system integration diagram.}
\label{fig.ChopperIntegration}
\end{figure}

The integration of the ESS Chopper Prototype at the V20 Test Instrument is shown in Fig.~\ref{fig.ChopperIntegration}.
Unlike the Source Pulse Choppers, the ESS Chopper Prototype not only communicates with the timing system, but also CHIC which is a controller unit for translating chopper manufacturer's specific data to a standardized common language for EPICS communication.
The ESS Chopper Prototype controller is based on a Beckhoff PLC (Programmable Logic Controller \cite{PLC}) and divided into two cpu cores. One core is dedicated to controlling the phase and the speed of the chopper, and one core is dedicated for communications between the CHIC and EPICS layer using Modbus TCP/IP \cite{TCPIP}. The ESS Chopper Prototype drive terminal is an EL7211 and the motor is a Beckhoff AM8121.
CHIC also verifies the correct chopper function through condition monitoring (motor temperature, voltage, current, vibration and overspeed are monitored). 

\subsection{EPICS and timing system}
The EPICS framework accesses the lower level controllers for individual hardware components at ESS \cite{Brodrick:2018mog,Pettersson:2018oto}. The ESS EPICS Environment (EEE) is a deployment system including a synchronised repository of EPICS modules across ESS and its in-kind partners.
Recalling Fig.~\ref{fig.NICOS}, both choppers are connected to the chopper drives. In the case of the ESS Chopper Prototype the drive is controlled from EPICS through CHIC.  EPICS commands such as start, stop and polling of data are sent from EPICS to the CHIC which reads the values from the ESS Chopper Prototype drive and sends back the data to EPICS.
Both chopper drives are controlled by the timing system reference signal while at the same time each chopper disc revolution is time stamped using a top dead center sensor \cite{TopDeadCenter} in the form of the TTL signal\cite{TTL}. The timing system then provides and gathers information from EPICS.
An EPICS forwarder (separate component) \cite{Mukai_2018} delivers information to Apache Kafka \cite{kafka}, a software platform used for building real-time data pipelines and streaming applications. Kafka is then directly providing information to the NeXus file writer, which saves the data to disk \cite{5.3}. 

\subsection{Detector system} 
The BMs used in the experiment are Multi Wire Proportional Chambers (MWPC) filled with $^3$He gas. The Front monitor is produced by the Mirrotron company \cite{Mirrotron} and the Back Monitor by ORDELA \cite{ORDELA}  with an efficiency of ca. $10^{-4}$ \cite{CharacterizationBeamMonitors}.
Both monitors were connected to ORTEC 142PC preamplifiers and  the output signal was connected to an 855 Dual Spectroscopy amplifier with 1 $\mu$s shaping time. 
The amplified signal of the Front Monitor, Back Monitor and the timing system was connected to the Data Acquisition System (DAQ) described in Fig.~\ref{fig.NICOS} as the Control and Timing demonstrator\cite{5.7}. The DAQ is a four channel system containing a Field Programmable Gate Array \cite{FPGA} unit (Xilinx Kintex Ultrascale KU040) on a commercially available Avnet Development Board \cite{FPGABoard}. The ADC (Analog to Digital converter) used for the DAQ is a Linear Technologies 2174 chip on a commercially-produced Open Hardware Design FMC-ADC-100m14b4cha board \cite{ADCchipConverter}.
The DAQ is a four channel system, that digitizes the output from the shaper. The Integrated Controls System Division (ICS) developed a centralized absolute timing system which uses a so-called master clock to distribute the absolute time to a number of timing receivers. In this test, a prescalar of 2 is applied to the 88.0525~MHz ICS master signal. The samples are timestamped by a local counter which is also synchronous to the ICS master. The resulting data is packaged into standard UDP datagrams and forwarded to the Event Formation Unit (EFU) \cite{Christensen_2018,5.6, 5.6_article} via a 1 Gb Ethernet link. 
\subsection{DMSC}
As shown in Fig.~\ref{fig.Architecture} the EFU is responsible for converting the detector data into neutron events. Slow control of the DAQ system is achieved via a serial interface to the EPICS system. The EFU forwards the data to Kafka, from which the file writer obtains the information to be written.
\subsection{V20 Test Instrument configuration}
\begin{figure}[h]
\onefigure[width=6.5cm]{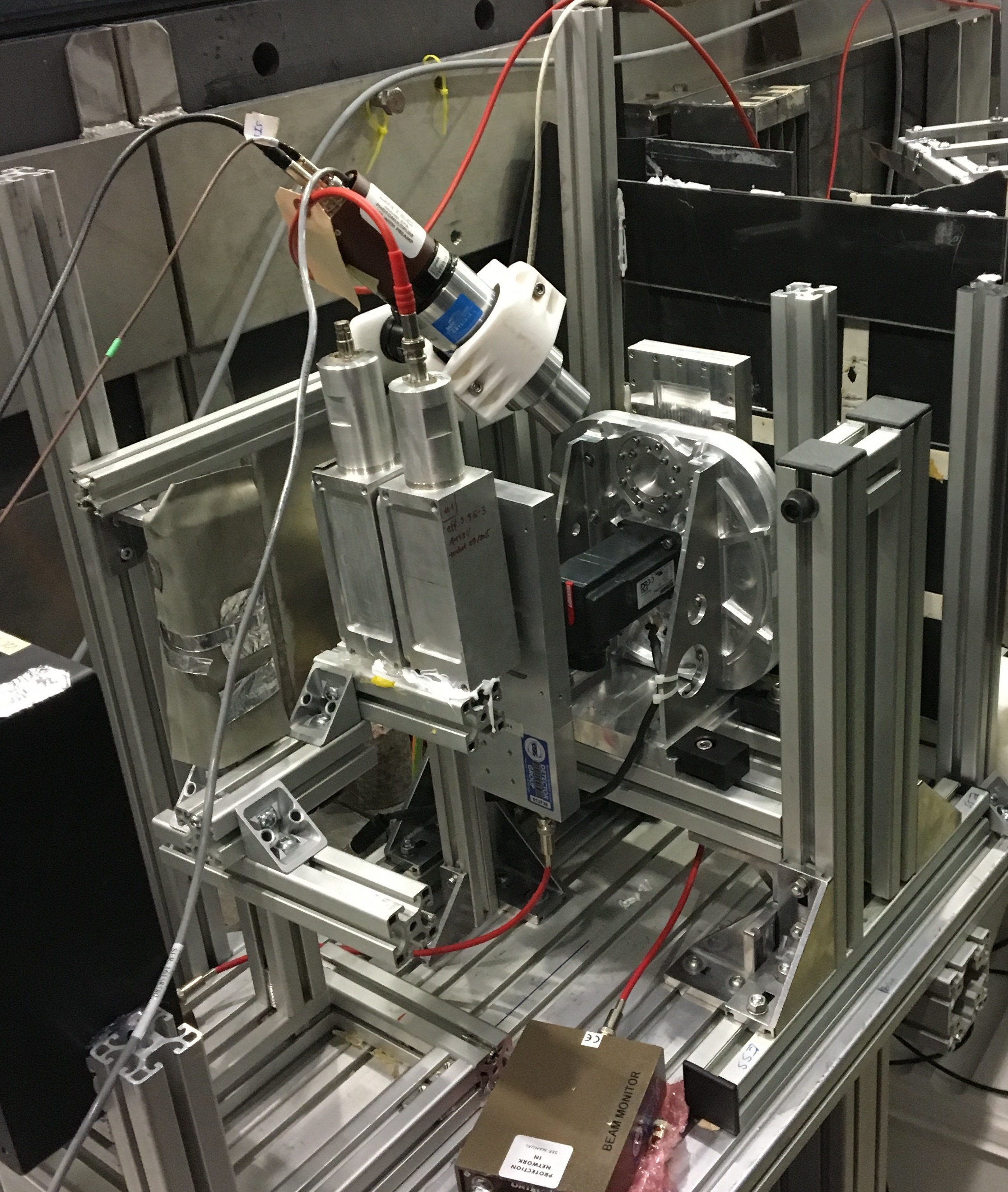}
\caption{The V20 Test Instrument was setup in native ESS pulse mode with the test stand located at approximately 32 m from the Source Pulse Choppers.}
\label{fig:V20}
\end{figure}

The V20 Test Instrument \cite{WORACEK} at HZB is designed to mimic the ESS pulse structure by using a counter-rotating double chopper system. These Source Pulse Choppers are installed at a distance of 21.75~m from the cold source. The ESS Chopper Prototype together with BMs (see Fig.~\ref{fig:V20}) were installed at a distance of approximately 32~m from the Source Pulse Choppers, i.e. 53.75~m from the cold source.
The Source Pulse Choppers are calibrated to provide the 2.86 ms pulse length with a repetition rate of 14 Hz. The integrated neutron flux at the V20 Test Instrument in this test configuration was 3~$\times$~10$^6$~n/cm$^2$/s. The beam was collimated so that the size of the beam incident to the ESS Chopper Prototype was 20~mm~$\times$~10~mm (of height and width, respectively) and therefore contained within the chopper window. The alignment of the whole setup was achieved via a laser system. The verification of the beam dimensions before and after the chopper was performed with a neutron camera\cite{NeutronOptics}. 

\section{Results}
\begin{figure}[h]
\onefigure[width=8.5cm]{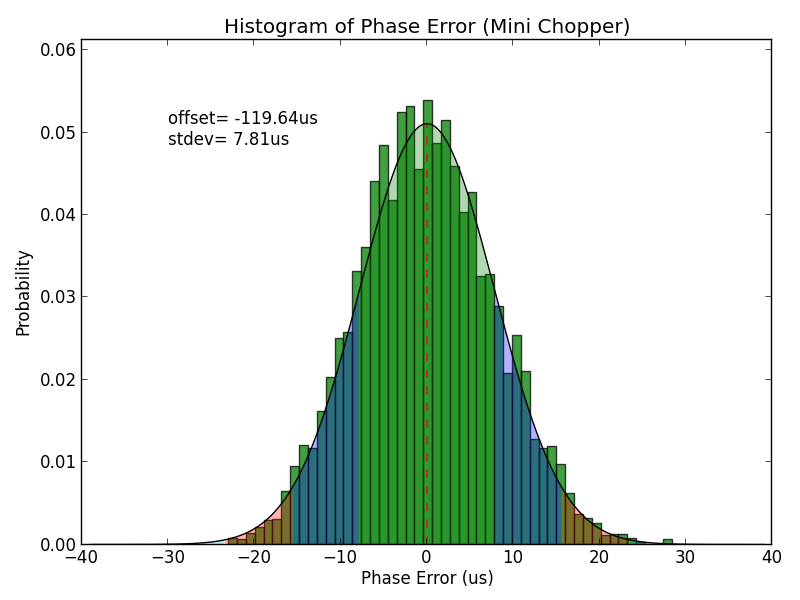}
\caption{Phase error histogram the integral of which is normalized to 1. The stdev is the standard deviation of the Gaussian distribution. The offset is an arbitrary number.} 
\label{fig.PhasekJitterHistogram}
\end{figure}

First the phase error of the ESS Chopper Prototype was monitored in real time using the EPICS Input/Output Controller.
In order to validate the functionality of the ESS Chopper Prototype phase, the phase error with respect to the main Source Pulse Choppers was recorded and plotted as seen in Fig.~\ref{fig.PhasekJitterHistogram}. As shown in the figure the  full width at the half maxima (FWHM) of the phase error is approximately 20~$\mu$s. This value is well within the required phase error for such a chopper.  

\begin{figure*}[h]
\begin{center}
\includegraphics[width=\textwidth]{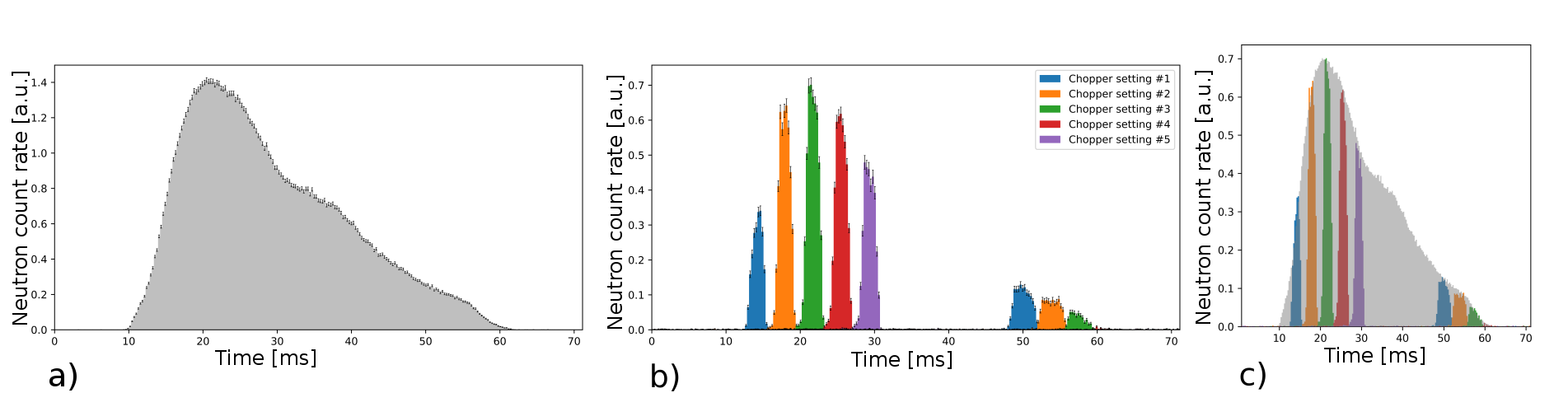}
\caption{Measured count rate distribution vs time at the: a)~front monitor just before entering the ESS Chopper Prototype, b)~back monitor and c)~monitor overlay. }
\label{fig.AllTogether}
\end{center}
\end{figure*}

While Fig.~\ref{fig.ChopperFunction} shows a cartoon of the evolution of the  distribution of neutrons as a function of time collected within one frame of the ESS period, Fig.~\ref{fig.AllTogether}a shows the experimental counterpart of Fig.~\ref{fig.ChopperFunction}c.  The distribution resembles the expected Boltzmann distribution of the cold moderator at HZB convoluted by instrumental effects (guide transmission function and attenuation). The peak observed at around 20 ms is the maximum neutron count rate of the V20 Test Instrument at ca 2.5~\r{A} and the data agree well with previously published data \cite{WORACEK}.
The collected data in the time-of-flight spectrum were aligned with known reference data recorded at the V20 Test Instrument. 
When compared to previous published data from the V20 Test Instrument at HZB \cite{WORACEK}, the ESS Chopper Prototype phase accurately followed the reference ESS source pulse. 
Fig.~\ref{fig.AllTogether}b (experimental counterpart of Fig.~\ref{fig.ChopperFunction}d) shows the neutrons after being modified by the ESS Chopper Prototype. The figure shows a collection of data sets from five different chopper phase settings indicated with different colours. The first data set was recorded using a chopper setting that had an initial phase aligned with known reference data recorded at the V20 Test Instrument \cite{WORACEK} with respect to the ESS source pulse. Subsequent data sets were recorded by using settings that were increased by a phase offset of 5 ms. Since the chopper has two openings, two transmitted pulses per ESS period per chopper setting were observed. The width of each of the produced pulses is a convolution of the chopper opening time and the dimensions of the beam before the chopper. Due to the fact that one opening is two times larger than the other opening, one can observe that the full width at half maximum of the wide pulses (chopper settings 1-3) are approximately two times as large as the narrow pulses of the same settings. The ratio is not exactly a factor two because of the finite size of the neutron beam. 
The reason that chopper setting 4 and 5 produce only 1 pulse is because the 180 degree pulse counter part is in the location of the phase space in which there are no neutrons. 
\par
In order to validate that the ESS Chopper Prototype functionality was as expected, the outgoing beam of the chopper was normalized with the incident beam to examine the distribution of the transmitted neutron frames. The respective peak intensity of the wavelength frames transmitted by the ESS Chopper Prototype corresponds satisfactorily to the incident beam (Fig.~\ref{fig.AllTogether}c).
\par
\begin{figure}[H]
\onefigure[width=8.5cm]{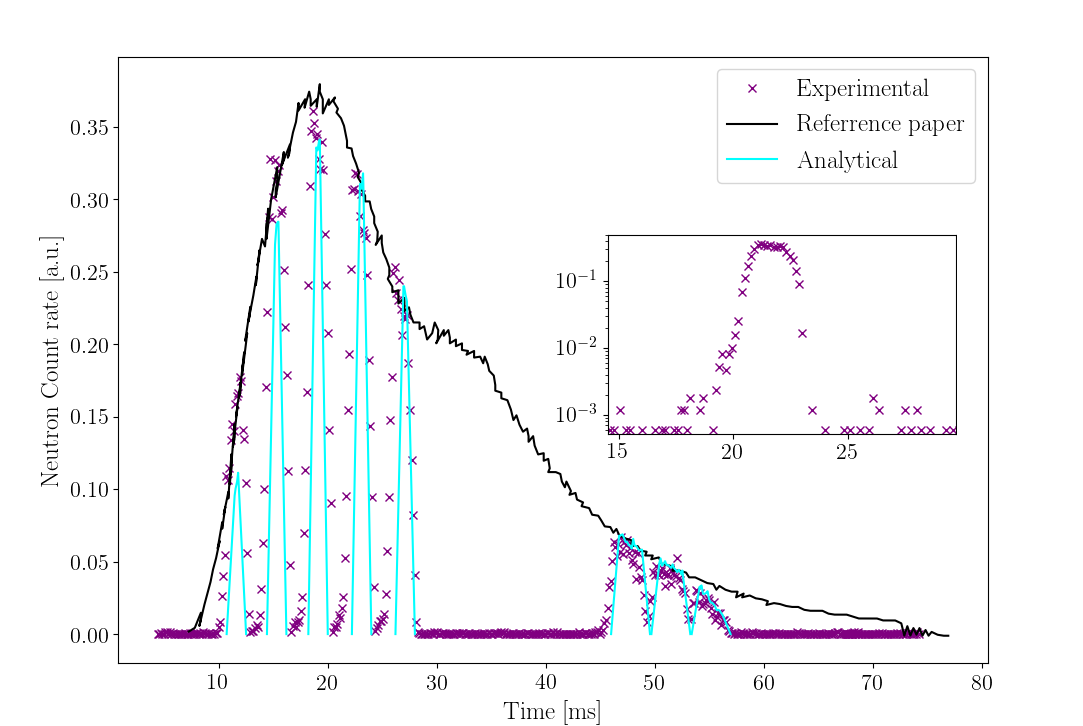}
\caption{Overlay of the experimentally measured neutron count rate before the ESS Chopper Prototype, analytically calculated chopper transmission normalized to the measured neutron count rate and the reference paper data expected before the ESS Chopper Prototype. The inset shows the neutron count rate of the third chopper opening (the third peak on the main panel).} 
\label{fig.TheoryAna}
\end{figure}

As a further diagnostic tool and to check the overall accuracy of our test architecture, the time-of-flight spectra were simulated using analytically calculated opening and closing chopper times (see Fig.~\ref{fig.TheoryAna}). This was convoluted with the incident time-of-flight (before the chopper) resulting in a good agreement between simulated and experimental data within the statistical error.
\par
The neutron absorption of the boron coating was checked by calculating the ratio of the peak transmission to the transmission of the chopper in its closed position (see inset of Fig.~\ref{fig.TheoryAna}). The measurement is statistically limited due to the low neutron counts. The absorption coefficient is close to 100\%. 
\section{Conclusion and Outlook}
Neutron data were successfully collected and time stamped using the proposed ESS neutron beam monitors and the ESS prototype detector readout electronics. The detector data were converted to neutron time-of-flight and analyzed. The test resulted in the expected TOF response before and after the ESS Chopper Prototype and compared excellently with instrument reference data, underlying the succesful interplay of all utilized hardware and software. Moreover the results of these tests demonstrate the diagnostic power of the testing procedure and the need for realistic control integration platforms. The results of this integration test validate the ESS chopper and detector DAQ architecture choices. Finally, the test was fundamental in helping the groups work together in different stages of instrument planning, integration and commissioning in preparation for ESS operations. 

\acknowledgments
We acknowledge the support of Helmholtz-Zentrum Berlin for supporting the beam time at the V20 Test Instrument. Furthermore we would like to acknowledge the support from BrightnESS, EU-H2020 grant number: 676548 for personnel and provision of monitors, readout prototypes, streaming, file writing system and others. Please note that David Broderick\textquotesingle s current affiliation is The Australian
National University, Canberra, ACT 2600 Australia. \nocite{ICALEPS_2015} 

\bibliographystyle{eplbib.bst}
\bibliography{references}

\begin{thebibliography}{10}
\expandafter\ifx\csname url\endcsname\relax\def\url#1{\texttt{#1}}\fi

\bibitem{TDRbook}
\Name{Peggs S. \etal} \Book{European Spallation Source Technical Design Report
  ESS-2013-001} 2013.
\newline\url{http://eval.esss.lu.se/cgi-bin/public/DocDB/ShowDocument?docid=274}

\bibitem{TechnicalDesignReport}
\Name{Peggs S.} \Book{{Technical Design of the ESS Facility}} presented at
  \Book{{4th International Particle Accelerator Conference (IPAC 2013):
  Shanghai, China, May 12-17, 2013}} 2013.
\newline\url{http://JACoW.org/IPAC2013/papers/thpwo074.pdf}

\bibitem{Garoby_2018}
\Name{Garoby R. \etal} \REVIEW{Physica Scripta}{93}{2018}{129501}.
\newline\url{https://doi.org/10.1088\%2F1402-4896\%2Faaecea}

\bibitem{First_choppers}
\Name{Dunning J.~R., Pegram G.~B., Fink G.~A., Mitchell D.~P. \and Segr\`e E.}
  \REVIEW{Phys. Rev.}{48}{1935}{704}.
\newline\url{https://link.aps.org/doi/10.1103/PhysRev.48.704}

\bibitem{Knoll}
\Name{Knoll G.~F.} \Book{Radiation detection and measurement} 2nd Edition
  (Wiley New York) 1989.
\newline\url{https://nla.gov.au/nla.cat-vn179294}

\bibitem{multiblade}
\Name{Piscitelli F., Messi F., Anastasopoulos M., Bry{\'{s}} T., Chicken F.,
  Dian E., Fuzi J., H{\"o}glund C., Kiss G., Orban J., Pazmandi P., Robinson
  L., Rosta L., Schmidt S., Varga D., Zsiros T. \and Hall-Wilton R.}
  \REVIEW{Journal of Instrumentation}{12}{2017}{P03013}.
\newline\url{https://doi.org/10.1088\%2F1748-0221\%2F12\%2F03\%2Fp03013}

\bibitem{multigrid}
\Name{Anastasopoulos M., Bebb R., Berry K., Birch J., Bry{\'{s}} T., Buffet
  J.-C., Clergeau J.-F., Deen P., Ehlers G., van Esch P., Everett S., Guerard
  B., Hall-Wilton R., Herwig K., Hultman L., H{\"o}glund C., Iruretagoiena I.,
  Issa F., Jensen J., Khaplanov A., Kirstein O., Higuera I.~L., Piscitelli F.,
  Robinson L., Schmidt S. \and Stefanescu I.} \REVIEW{Journal of
  Instrumentation}{12}{2017}{P04030}.
\newline\url{https://doi.org/10.1088\%2F1748-0221\%2F12\%2F04\%2Fp04030}

\bibitem{gdgem}
\Name{Pfeiffer D., Resnati F., Birch J., Etxegarai M., Hall-Wilton R.,
  H{\"o}glund C., Hultman L., Llamas-Jansa I., Oliveri E., Oksanen E., Robinson
  L., Ropelewski L., Schmidt S., Streli C. \and Thuiner P.} \REVIEW{Journal of
  Instrumentation}{11}{2016}{P05011}.
\newline\url{https://doi.org/10.1088\%2F1748-0221\%2F11\%2F05\%2Fp05011}

\bibitem{MaPMT}
\Name{Rofors E., Perrey H., Jebali R.~A., Annand J., Boyd L., Clemens U.,
  Desert S., Engels R., Fissum K., Frielinghaus H., Gheorghe C., Hall-Wilton
  R., Jaksch S., Jalgén A., Kanaki K., Kemmerling G., Maulerova V., Mauritzson
  N., Montgomery R., Scherzinger J. \and Seitz B.} \REVIEW{Nuclear Instruments
  and Methods in Physics Research Section A: Accelerators, Spectrometers,
  Detectors and Associated Equipment}{}{2019}{}.
\newline\url{http://www.sciencedirect.com/science/article/pii/S0168900219303134}

\bibitem{CDT}
\Name{{CDT GmbH}} \Book{Cascade detector technologies} (2019).
\newline\url{http://n-cdt.com/}

\bibitem{Kirstein2014}
\Name{Kirstein O. \etal} \Book{{Neutron Position Sensitive Detectors for the
  ESS}} presented at \Book{{23rd International Workshop on Vertex Detectors
  (Vertex 2014): Doksy, Czech Republic, September 15-19, 2014}}.
\newline\url{{https://arxiv.org/abs/1411.6194}}

\bibitem{HENSKE2012151}
\Name{Henske M., Klein M., Köhli M., Lennert P., Modzel G., Schmidt C. \and
  Schmidt U.} \REVIEW{Nuclear Instruments and Methods in Physics Research
  Section A: Accelerators, Spectrometers, Detectors and Associated
  Equipment}{686}{2012}{151 }.
\newline\url{http://www.sciencedirect.com/science/article/pii/S016890021200589X}

\bibitem{CharacterizationBeamMonitors}
\Name{Issa F., Khaplanov A., Hall-Wilton R., Llamas I., Riktor M.~D., Brattheim
  S.~R. \and Perrey H.} \REVIEW{Phys. Rev. Accel. Beams}{20}{2017}{092801}.
\newline\url{https://link.aps.org/doi/10.1103/PhysRevAccelBeams.20.092801}

\bibitem{ORDELA}
\Name{ORDELA} \Book{Ordela inc.} (2019).
\newline\url{http://www.ordela.com/}

\bibitem{Mirrotron}
\Name{Mirrotron} \Book{Mirrotron} (2019).
\newline\url{http://www.mirrotron.kfkipark.hu/}

\bibitem{4.1}
\Name{{Kolya, S. \etal}} \Book{Integration plan for detector readout} (2017).
\newline\url{https://doi.org/10.17199/BRIGHTNESS.D4.1}

\bibitem{TOForigin}
\Name{Wolff M.~M. \and Stephens W.~E.} \REVIEW{Review of Scientific
  Instruments}{24}{1953}{616}.
\newline\url{https://doi.org/10.1063/1.1770801}

\bibitem{EPICS}
\Name{Dalesio L.~R., Hill J.~O., Kraimer M., Lewis S., Murray D., Hunt S.,
  Watson W., Clausen M. \and Dalesio J.} \REVIEW{Nuclear Instruments and
  Methods in Physics Research Section A: Accelerators, Spectrometers, Detectors
  and Associated Equipment}{352}{1994}{179 }.
\newline\url{http://www.sciencedirect.com/science/article/pii/0168900294914931}

\bibitem{ESSVerticalIntegration}
\Name{{Gahl} T., {Hagen} M., {Hall-Wilton} R., {Kolya} S., {K{\"o}nnecke} M.,
  {Rescic} M., {Rod} T.~H., {Sutton} I., {Trahern} G. \and {Kirstein} O.}
  \Book{{Hardware Aspects, Modularity and Integration of an Event Mode Data
  Acquisition and Instrument Control for the European Spallation Source (ESS)}}
  presented at \Book{{Proc. ICANS XXI, Mito, Japan, JAEA-Conf 2015-002
  (2014)}}.
\newline\url{https://arxiv.org/abs/1507.01838}

\bibitem{Mukai_2018}
\Name{Mukai A., Clarke M., Christensen M., Nilsson J., Shetty M., Brambilla M.,
  Werder D., K{\"o}nnecke M., Harper J., Jones M., Akeroyd F., Reis C.,
  Kourousias G. \and Richter T.} \REVIEW{Journal of
  Instrumentation}{13}{2018}{T10001}.
\newline\url{https://doi.org/10.1088\%2F1748-0221\%2F13\%2F10\%2Ft10001}

\bibitem{Christensen_2018}
\Name{Christensen M., Shetty M., Nilsson J., Mukai A., Jebali R.~A., Khaplanov
  A., Lupberger M., Messi F., Pfeiffer D., Piscitelli F., Blum T., S{\o}gaard
  C., Skelboe S., Hall-Wilton R. \and Richter T.} \REVIEW{Journal of
  Instrumentation}{13}{2018}{T11002}.
\newline\url{https://doi.org/10.1088\%2F1748-0221\%2F13\%2F11\%2Ft11002}

\bibitem{Mantid}
\Name{Arnold O., Bilheux J., Borreguero J., Buts A., Campbell S., Chapon L.,
  Doucet M., Draper N., Leal R.~F., Gigg M., Lynch V., Markvardsen A.,
  Mikkelson D., Mikkelson R., Miller R., Palmen K., Parker P., Passos G.,
  Perring T., Peterson P., Ren S., Reuter M., Savici A., Taylor J., Taylor R.,
  Tolchenov R., Zhou W. \and Zikovsky J.} \REVIEW{Nuclear Instruments and
  Methods in Physics Research Section A: Accelerators, Spectrometers, Detectors
  and Associated Equipment}{764}{2014}{156 }.
\newline\url{http://www.sciencedirect.com/science/article/pii/S0168900214008729}

\bibitem{CHIC}
\Name{ESS} \Book{Choppers research and development} (2019).
\newline\url{https://europeanspallationsource.se/instrument-technologies/chopper-systems/choppers-research-development\#chic}

\bibitem{WORACEK}
\Name{Woracek R., Hofmann T., Bulat M., Sales M., Habicht K., Andersen K. \and
  Strobl M.} \REVIEW{Nuclear Instruments and Methods in Physics Research
  Section A: Accelerators, Spectrometers, Detectors and Associated
  Equipment}{839}{2016}{102 }.
\newline\url{http://www.sciencedirect.com/science/article/pii/S0168900216309597}

\bibitem{STROBL201374}
\Name{Strobl M., Bulat M. \and Habicht K.} \REVIEW{Nuclear Instruments and
  Methods in Physics Research Section A: Accelerators, Spectrometers, Detectors
  and Associated Equipment}{705}{2013}{74 }.
\newline\url{http://www.sciencedirect.com/science/article/pii/S0168900212016142}

\bibitem{FORSTER2018298}
\Name{F{\"o}rster D.~F., M{\"u}ller F., Giesen U., Lindenau B., Ortmanns T.,
  Wolters J., Pabst U., Butzek M., Woracek R., Kozielewski T. \and Monkenbusch
  M.} \REVIEW{Nuclear Instruments and Methods in Physics Research Section A:
  Accelerators, Spectrometers, Detectors and Associated
  Equipment}{908}{2018}{298 }.
\newline\url{http://www.sciencedirect.com/science/article/pii/S0168900218308866}

\bibitem{MasterCam}
\Name{MasterCam} \Book{Mastercam software} (2019).
\newline\url{https://www.mastercam.com/en-us/}

\bibitem{PLC}
\Name{Hanssen D.} \Book{Programmable Logic Controllers: A Practical Approach to
  IEC 61131-3 using CoDeSys} (Wiley) 2015.
\newline\url{https://books.google.se/books?id=My-PCgAAQBAJ}

\bibitem{TCPIP}
\Name{Kozierok C.~M.} \Book{{The TCP/IP guide: a comprehensive, illustrated
  internet protocols reference}} (No Starch Press, San Francisco, CA) 2005 the
  book can be consulted by contacting: PH-CMD: Bukowiec, Sebastian Czeslaw.
\newline\url{https://cds.cern.ch/record/1249759}

\bibitem{Brodrick:2018mog}
\Name{Brodrick D., Brys T., Korhonen T. \and Sparger J.} \Book{{EPICS
  Architecture for Neutron Instrument Control at the European Spallation
  Source}} presented at \Book{{16th International Conference on Accelerator and
  Large Experimental Physics Control Systems (ICALEPCS 2017): Barcelona, Spain,
  October 8-13, 2017}}.
\newline\url{{http://accelconf.web.cern.ch/accelconf/icalepcs2017/papers/webpl01.pdf}}

\bibitem{Pettersson:2018oto}
\Name{Pettersson A., Brodrick D., Brys T. \and Hartl M.} \Book{{Integration of
  Sample Environment Systems at ESS}} presented at \Book{{16th International
  Conference on Accelerator and Large Experimental Physics Control Systems
  (ICALEPCS 2017): Barcelona, Spain, October 8-13, 2017}}.
\newline\url{{http://accelconf.web.cern.ch/AccelConf/icalepcs2017/talks/webpl01_talk.pdf}}

\bibitem{TopDeadCenter}
\Name{{Rose} C.~R., {Lara} P.~D. \and {Nelson} R.~O.} \Book{A simple method for
  real-time dsp-based neutron chopper speed and phase control} in proc. of
  \Book{PACS2001. Proceedings of the 2001 Particle Accelerator Conference (Cat.
  No.01CH37268)} Vol.~2 2001 pp. 1444--1446 vol.2.
\newline\url{{https://permalink.lanl.gov/object/tr?what=info:lanl-repo/lareport/LA-UR-01-3301}}

\bibitem{TTL}
\Name{Wikipedia} \Book{Transistor\textemdash transistor logic, wikipedia, the
  free encyclopedia} online; accessed 05-April-2019 (2019).
\newline\url{https://en.wikipedia.org/wiki/Transistor\%E2\%80\%93transistor\_logic}

\bibitem{kafka}
\Name{Garg N.} \Book{Apache Kafka} (Packt Publishing) 2013.
\newline\url{{ftp://xil80.duckdns.org/upload/books/BigData/Learning\%20Apache\%20Kafka,\%202nd\%20Edition\%20Start\%20from\%20scratch\%20and\%20learn\%20how\%20to\%20administer\%20Apache\%20Kafka\%20effectively\%20for\%20messaging.pdf}}

\bibitem{5.3}
\Name{{Mukai, A. \etal}} \Book{Beta-version data aggregator software} (2018).
\newline\url{https://doi.org/10.17199/BRIGHTNESS.D5.3}

\bibitem{5.7}
\Name{{Nilsson, J. \etal}} \Book{Software fast field acquisition} (2018).
\newline\url{https://doi.org/10.17199/brightness.d5.7}

\bibitem{FPGA}
\Name{Kuon I., Tessier R. \and Rose J.} \REVIEW{Found. Trends Electron. Des.
  Autom.}{2}{2008}{135}.
\newline\url{http://www.eecg.toronto.edu/~jayar/pubs/kuon/foundtrend08.pdf}

\bibitem{FPGABoard}
Avnet Electronics Marketing \Book{Kintex UltraScale KU040 Development Board
  Hardware User guide}.
\newline\url{https://www.avnet.com/opasdata/d120001/medias/docus/13/aes-AES-KU040-DB-G-User-Guide.pdf}

\bibitem{ADCchipConverter}
\Name{Firmiarz M., Cattin M. \and van~der Bij E.} \Book{Fmcadc100m14b4cha
  board} (2019).
\newline\url{https://www.ohwr.org/project/fmc-adc-100m14b4cha/wikis/home}

\bibitem{5.6}
\Name{Christensen M.~J., Richter T., Shetty M., Nilsson J., Skelboe S. \and
  S{\o}gaard C.} \Book{Software neutron event data processing} (2018).
\newline\url{https://doi.org/10.17199/brightness.d5.6}

\bibitem{5.6_article}
\Name{Christensen M., Shetty M., Nilsson J., Mukai A., Jebali R.~A., Khaplanov
  A., Lupberger M., Messi F., Pfeiffer D., Piscitelli F., Blum T., S{\o}gaard
  C., Skelboe S., Hall-Wilton R. \and Richter T.} \REVIEW{Journal of
  Instrumentation}{13}{2018}{T11002}.
\newline\url{https://doi.org/10.1088\%2F1748-0221\%2F13\%2F11\%2Ft11002}

\bibitem{NeutronOptics}
\Name{{Neutron Optics, GRENOBLE}} \Book{Neutron optics} (2019).
\newline\url{http://neutronoptics.com/}

\bibitem{ICALEPS_2015}
\Name{Gahl T., Hall-Wilton R., Kirstein O., Korhonen T., Richter T.,
  Sandstr{\"o}m A., Sutton I. \and Taylor J.} \Book{{The Modular Control
  Concept of the Neutron Scattering Experiments at the European Spallation
  Source ESS}} presented at \Book{{Proceedings, 15th International Conference
  on Accelerator and Large Experimental Physics Control Systems (ICALEPCS
  2015): Melbourne, Australia, October 17-23, 2015}}.
\newline\url{http://icalepcs.synchrotron.org.au/html/auth1043.htm}

\end{thebibliography}
\end{document}